# A THEORETICAL LOOK AT ORDINAL CLASSIFICATION METHODS BASED ON COMPARING ACTIONS WITH LIMITING BOUNDARIES BETWEEN ADJACENT CLASSES

**Eduardo Fernández**
Universidad Autónoma de Coahuila, México (eddyf171051@gmail.com)

**José Rui Figueira**
CEG-IST, Instituto Superior Técnico, Universidade de Lisboa, Portugal ( figueira@tecnico.ulisboa.pt)

**Jorge Navarro**
Universidad Autónoma de Sinaloa, México
(jnavarro@uas.edu.mx)

**Abstract.**

This paper addresses the general problem of designing ordinal classification methods based on comparing actions with limiting boundaries of ordered classes (categories). The fundamental requirement of the method consists of setting a relational system $(D,S)$, where $S$ and $D$ are reflexive and transitive relations, respectively, $S$ should be compatible with the order of the set of classes, and $D$ is a subset of $S$. An asymmetric preference relation $P$ is defined from $S$. Other requirements are imposed on the actions which compose the limiting boundaries between adjacent classes, in such a way that each class is closed below and above. The paper proposes $S$-based and $P$-based assignment procedures. Each of them is composed of two complementary assignment procedures, which correspond through the transposition operation and should be used conjointly. The methods work under several basic conditions on the set of limiting boundaries. Under other more demanding separability requirements, each procedure fulfills the set of structural properties established for other outranking-based ordinal classification methods. Our proposal avoids the conflict between the required correspondence through the transposition operation and the assignment of limiting actions to the classes to which they belong. We thus propose very diverse $S$ and $P$-based ordinal classification approaches with desirable properties, which can be designed by using decision models with the capacity to build preference relations fulfilling the basic requirements to $S$ and $D$.

# 1. Introduction

Ordinal classification is a main topic in the field of multiple criteria decision-making (MCDM), which has attracted attention from the MCDM community over the last decades. Unlike nominal classification, in ordinal classification objects, alternatives, actions are assigned to ordered and pre-defined classes or categories.

Many multi-criteria ordinal classification methods have been proposed in the MCDM literature. The main differences among them are focused on: a) the underlying preference model; and b) the way in which classes are characterized.

Regarding the decision-maker's (DM's) preference model, a vast majority of the methods comes from one of three main paradigms:

- The functional paradigm, based on the construction of value functions (e.g., Jacquet-Lagrèze, 1995; Zopounidis and Doumpos, 2000);
- Symbolic methods connected with Artificial Intelligence (e.g., Greco *et al*., 2001); and
- The relational paradigm (e.g., Massaglia and Ostanello, 1991; Yu, 1992).

Whatever the model of preferences, classes should be characterized in some way. There are two main ways:

  i)     Using limiting actions as boundaries between adjacent classes (e.g. Yu, 1992; Roy and Bouyssou, 1993; Perny, 1998; Araz and Ozkarahan, 2007; Nemery and Lamboray, 2008; Ishizaka et al., 2012);

  ii)    Using decision examples (or reference actions), which are representative of the related classes, and whose classification is (or maybe) known (e.g. Jacquet-Lagrèze, 1995; Zopounidis and Doumpos, 2000; Greco et al., 2001; Köksalan and Ulu, 2003; Almeida-Dias et al., 2010, 2012; Fernandez and Navarro. 2011).

Among the methods in Point i), the most popular one is is ELECTRE TRI, proposed by Yu (1992), detailed by Roy and Bouyssou (1993), and renamed by Almeida-Dias et al. (2010) as ELECTRE TRI-B (see Govindan and Jepsen, 2016, for a summary of published applications of this method).

In ELECTRE TRI-B, classes are characterized by introducing a single limiting action (profile) between adjacent classes, which are closed from below. In other words, for adjacent classes, the limiting action between them belongs to the upper category. ELECTRE TRI-B uses outranking and preference relations to compare the actions with the limiting profiles. Actions are assigned according to the results of this comparison. ELECTRE TRI-B consists of two alternative procedures, which are based on two different logics. The pseudo-conjunctive procedure uses an outranking relation, whereas the pseudo-disjunctive procedure exploits an asymmetric preference relation. Both assignment rules fulfil several fundamental properties (conformity, monotonicity, stability, homogeneity, independence, and uniqueness), which are a paradigm for outranking based ordinal classification methods. The procedures can be used alternatively, according to the logic with which the decision-maker (DM) feels more comfortable. Both procedures could be used conjointly, although there are no theoretical arguments to justify it. The pseudo-conjunctive procedure has received more attention than the other one. The pseudo-conjunctive method has been axiomatically addressed by Bouyssou and Marchant (2007) and most of the ELECTRE TRI-B applications have been performed by using this procedure.

The original ELECTRE TRI evolved in different directions:

- Almeida-Dias et al. (2010, 2012) proposed the ELECTRE TRI-C and TRI-nC methods, in which categories are described by representative or characteristic actions. If $R_k$ denotes the set of representative actions of class $C_k$, in order to assign an action $x$, ELECTRE TRI-nC (TRI-C is a particular case) exploits an outranking relation $S$ between actions and representative sets of classes. The method uses a descending (respectively ascending) procedure, based on finding the first $C_k$ for which $xSR_k$ (resp. $R_kSx$). Due to the symmetry between both rules, Almeida-Dias et al. (2010, 2012) suggest that they should be used conjointly, and the assignment may be an interval of classes. Such a symmetry has a deep meaning: Almeida-Dias et al. (2010) argued that the ascending and descending procedures correspond via the transposition operation. This operation consists of inverting the direction of preferences on all criteria, also inverting the ordering of the classes. According to Bouyssou and Marchant (2015), the conclusions obtained after this operation should not be different from the original conclusions. It should be remarked that the assignment rules in ELECTRE TRI-B do not correspond via the transposition operation (Roy, 2002; Bouyssou and Marchant, 2015). This is a consequence of considering classes as closed from below (Bouyssou and Marchant, 2015).

- To enhance the definition of the limiting boundaries, ELECTRE TRI-nB was recently proposed (Fernández *et al*., 2017). In this enhanced method, limiting boundaries between adjacent (consecutive) classes are characterized through several or many limiting profiles. Compared to ELECTRE TRI-B, the ELECTRE TRI-nB uses a richer information about the relations between actions and limiting boundaries. As in the original method, in ELECTRE TRI-nB, the classes keep closed from below, both assignment rules fulfil the same set of structural properties, and the procedures have no symmetry with respect to the transposition operation.

- ELECTRE TRI-nB was extended to the interval framework by Fernández et al. (2020). In INTERCLASS-nB, criterion performances levels, weights, veto and majority thresholds may be interval numbers, what permits an easy handle of imprecision, uncertainty, ill-definition and arbitrariness. Other characteristics and properties are similar to ELECTRE TRI-B and TRI-nB (Fernández *et al.*, 2020).
- A hierarchical ELECTRE TRI-B with interacting criteria was proposed by Corrente *et al.* (2016). Ordinal classification problems can be solved in different levels of the hierarchy. The fulfilment of the structural properties (may be altered by interaction of criteria) was not addressed by this paper.
- Bouyssou and Marchant (2015) proposed an interesting and important ascending procedure based on an outranking relation *S*, called the dual pseudo-conjunctive ELECTRE TRI-B. If $b_k$ denotes the limiting action between classes $C_k$ and $C_{k+1}$, the dual pseudo-conjunctive procedure bases the assignment of an action *x* on the fulfilment of $b_kSx$ instead of $xSb_k$, as in the original (renamed primal) pseudo-conjunctive procedure. Since both primal and dual procedures correspond via the transposition operation, Bouyssou and Marchant (2015) support their conjoint use (similarly to the ascending and descending rules in ELECTRE TRI-C). In the proposal from Bouyssou and Marchant (2015), limiting profiles are interpreted as fictitious actions in the frontier of adjacent classes, but belonging to no one. This could make very difficult a direct elicitation of such fictitious profiles. Bouyssou and Marchant (2015) suggested the use of indirect methods from the preference-disaggregation analysis to elicit the limiting actions. The profile $b_k$ is assigned to $C_{k+1}$ (respectively $C_k$) by the primal (resp. dual) procedure; hence, the conformity property is not fulfilled, and the fictitious character of the limiting actions is underlined.
- Bouyssou *et al.* (2020) perform a complete and very interesting axiomatic characterization of the pseudo-conjunctive ELECTRE TRI-nB. They proved that the the ELECTRE TRI-nB pseudo-conjunctive procedure is more general than the additive value function model for ordinal classification, and suggest the use of a conjoint primal-dual pseudo-conjunctive approach having several or many limiting actions between consecutive classes. The limiting actions do not belong to any class as in Bouyssou and Marchant (2015).

Thus, after the paper by Bouyssou and Marchant (2015), a pair decision analyst-decision maker interested in using ELECTRE TRI-B or a variant, faces the following dilemma:

    i.      To use the pseudo-conjunctive procedure;

ii. To use the pseudo-disjunctive procedure:

iii. To choose a combined use of both procedures, and consider the interval of classes derived from this combination;

iv. To choose the conjoint primal-dual method by Bouyssou and Marchant (2015), in spite of the difficulty to set the limiting actions, and the non-fulfilment of the Conformity Property.

As stated before, most of the published applications have used the single pseudo-conjunctive procedure, may be for its popularity, for having a complete axiomatic foundation, or for disposing of simpler indirect elicitation methods than the pseudo-disjunctive procedure. However, the single use of the pseudo conjunctive procedure is asymmetric. The information provided by $xSb_k$ and $not(xSb_{k+1})$ has no more value than the obtained from $b_kSx$ and $not(b_{k-1}Sx)$. Then, why only $xSb_k$ and $not(xSb_{k+1})$ should be taken into account? The conjoint use of both kinds of information should produce more informed assignments. It is a central idea in support of the combined use of procedures with symmetry respect to the transposition operation. Nevertheless, the set of structural properties, including conformity, of ELECTRE TRI-B should be kept.

Roy (2002) and Bouyssou and Marchant (2015) identified a conflicting nature between the Conformity Property and the correspondence through the transposition operation. Roy (2002) prioritized the fulfillment of the Conformity Property, whereas Bouyssou and Marchant (2015) and Bouyssou et al. (2020) privileged the consistency with the transposition operation. In this paper, we suggest a way to avoid such a contradiction redefining the limiting boundaries between adjacent classes. We also attempt to generalize the main idea behind ELECTRE TRI-B, its variants, and other methods based on limiting profiles between adjacent classes.

Consider any relation "$x$ is at least as good as $y$ with respect to a certain desirable property $\Xi$", or, alternatively, any relation "$x$ is preferred to $y$ with respect to a certain desirable property $\Xi$"; there is no matter the way to create these relations. Suppose also that the pair DM-decision analyst wants to use one of these relations to design a limiting boundary-based assignment method to classes ordered in the sense of increasing $\Xi$. In this paper we address the following fundamental issue: To propose a general form of the decision rule, and to identify the requirements on the limiting actions to fulfill: i) the entire set of structural properties originally suggested by Roy and Bouyssou (1993); and ii) symmetry with respect to the transposition operation. This general characterization is inspired by the primal and dual rules proposed by Bouyssou and Marchant (2015), but it is more general and satisfies the Conformity Property. To achieve the compatibility with this property, we require to introduce two "layers" of actions to describe each limiting boundary. Such an addition becomes more demanding the requirements on the set of limiting actions than the ones in ELECTRE TRI-nB. Nevertheless, the methods still work under weaker requirements, although lose some of the properties.

The paper is structured as follows: an outranking-based primal and dual method is discussed in Section 2, including its properties. An asymmetric preference-based method is characterized in Section 3. Two examples illustrating the methods are presented in Section 4, followed by a discussion section in which the generality of the proposal is underlined. Lastly, some conclusions and lines of future research are presented in Section 6.

## 2. An outranking-based primal and dual conjoint procedure

This section introduces the basic requirements, the new method, and the structural properties it must fulfill.

### 2.1 Requirements

Some conditions and results on the relational system of preferences and the limiting actions must be established before the presentation of the method.

**Condition 1** (**Requirements on the relational system of preferences**)[1]

Let us consider a pair of binary relations ($D, S$) with the following characteristics:

- $D$ is a transitive relation;
- $xSy$ may be interpreted as "action $x$ has at least as much of certain desirable property $\Xi$ as action $y$". $S$ is a reflexive relation. From $S$, an asymmetric preference relation $P$ is defined as $xPy \Leftrightarrow xSy$ and $not(ySx)$;
- ($D,S$) must fulfill:

    $\forall\ (x, y, z) \in A \times A \times A$:

    i. $xDy \Rightarrow xSy$;

    ii. $xSy$ and $yDz \Rightarrow xSz$;

    iii. $xDy$ and $ySz \Rightarrow xSz$

**Proposition 1 (Properties of the combined use of $P$ and $D$):**

---

[1] Under very general conditions, "at least as good as" relations created by multi-criteria decision methods and the Pareto dominance relation fulfill this requirement.

Under Condition 1, the pair (*D, P*) satisfies the following properties:

i. $xPy$ and $yDz \Rightarrow xPz$;

ii. $xDy$ and $yPz \Rightarrow xPz$;

*Proof:*

Proposition 1.i: $xPy$ and $yDz \Rightarrow xSy$ and $yDz \Rightarrow xSz$ (Condition 1.ii); Suppose that $zSx$; then $yDz$ and $zSx \Rightarrow ySx$ (Condition 1.iii). This contradicts $xPy$; hence $not(zSx)$. Finally, $xSz$ and $not(zSx) \Leftrightarrow xPz$.

Proposition 1.ii: The proof follows the same logic as above, but using first Condition 1.iii.

Next, we present the requirements to the limiting profiles of an ordinal classification method based on a relational system which fulfills Condition 1.

**Condition 2 (Basic demands on limiting profiles)**

Consider a set of *M* ordered and predefined classes $C= C_1,\ldots,C_k,\ldots,C_M$ , ($M \geq 2$) (ordered in the sense of increasing $\Xi$). The boundary between $C_k$ and $C_{k+1}$ is described by a set of limiting actions $B_k$, for $k=1,\ldots,M$-1. Each $B_k$ is composed of two disjoint subsets $B_{Uk}$ and $B_{Lk}$ such that:

i. Each $w \in B_{Uk}$ is in $C_k$;

ii. Each $z \in B_{Lk}$ is in $C_{k+1}$;

iii. There is no pair $(w,z) \in B_{Lk}\ B_{Lk}$ fulfilling $wPz$;

iv. There is no pair $(w,z) \in B_{Uk}\ B_{Uk}$ fulfilling $wPz$;

**Condition 3 (Separability conditions)**

i. There is no pair $(w,z) \in B_{Uk}\ B_{Lk}$ fulfilling $wSz$;

ii. There is no pair $(w,z) \in B_k\ B_h$ ($h>k$) fulfilling $wSz$;

iii. For each $z \in B_{Uk}$ there is $y \in B_{Lk-1}$ such that $zSy$;

iv. For each $z \in B_{Lk}$ there is $y \in B_{Uk+1}$ such that $ySz$;

v. For each $z \in B_{Uk}$ there is $w \in B_{Uk+1}$ such that $wDz$;

vi. For each $z \in B_{Uk}$ there is $w \in B_{Uk-1}$ such that $zDw$;

vii. For each $z \in B_{Lk}$ there is $y \in B_{Lk-1}$ such that $zDy$;

viii. For each $z \in B_{Lk}$ there is $y \in B_{Lk+1}$ such that $yDz$

Condition 3.i is a separability condition between the subsets which compound each limiting boundary. 3.ii-viii are separability conditions between boundaries. Conditions 3.i-3.iv are *S*-based separability requirements; 3.v-3.viii are *D*-based separability conditions. Let us remark that Conditions 3.ii, 3.iii, and 3.iv arise naturally from the order of classes in the sense of increasing $\Xi$.

## 2.2 The *S*-based method

The method is presented in the next three definitions.

**Definition 1** (*S*- relation between actions and boundaries)

a) $xSB_k \Leftrightarrow$ There is $w \in B_{Lk}$ such that $xSw$ and there is no $z \in B_k$ fulfilling $zPx$;
b) $B_kSx \Leftrightarrow$ There is $w \in B_{Uk}$ such that $wSx$ and there is no $z \in B_k$ fulfilling $xPz$

**Definition 2** (*S*-based primal assignment procedure)

Set the relational system $(D,S)$ fulfilling Condition 1. Set $B_k$, $k=1,\ldots M-1$, fulfilling Condition 2. Assume that $B_0$ is the anti-ideal action, and $xSB_0$ for all $x$.

  i. For $k=M-1,\ldots,0$, find the first limiting boundary $B_k$ fulfilling $xSB_k$;
  ii. Take $C_{k+1}$ as an acceptable class to assign $x$.

**Definition 3** (*S*-based dual assignment procedure)

Set the relational system $(D,S)$ fulfilling Condition 1. Set $B_k$, $k=1,\ldots M-1$, fulfilling Condition 2. Assume that $B_M$ is the ideal action, and $B_MSx$ for all $x$.

  a. For $k=1,\ldots M$, find the first $B_k$ fulfilling $B_kSx$;
  b. Take $C_k$ as an acceptable class to assign $x$.

**Remark 1:**

Note that the primal and dual procedures correspond through the transposition operation; they should be used conjointly in assigning actions. Note also that if $B_{Uk}$ is empty ($k=1, \ldots M-1$), the primal rule under Condition 2.ii, iii, and Condition 3.ii, vii and viii is equivalent to the pseudo-conjunctive procedure of ELECTRE TRI-nB.

**Proposition 2 (Basic properties of the *S* relation between actions and limiting boundaries)**

Under Conditions 1, 2 and 3.v-3.viii, the following properties are fulfilled:

i) $xSB_k \Rightarrow xSB_h$ for $k>h$;
ii) $B_kSx \Rightarrow B_hSx$ for $h>k$

(See the proof in Appendix A).

## 2.3 Structural properties of the *S*-based procedures[2]

We should prove that the *S*-based primal and dual procedures satisfy the properties which were firstly proposed by Roy and Bouyssou (1993) for ELECTRE TRI-B. These properties are fulfilled also by ELECTRE TRI-nB (Fernández *et al*., 2017) and INTERCLASS-nB (Fernández *et al*., 2020), and, as stated in the introduction, are paradigmatic for outranking-based ordinal classification methods.

Let us recall after (Fernández *et al*., 2017, 2020) the definition of merging and splitting operations on the set of classes.

**Definition 4** (**Merging and splitting operations**)

a) Merging: A merging operations puts together two consecutive categories, $C_k$ and $C_{k+1}$, by forming a new category, denoted by $C'_k$. This operation consists of removing the set of limiting profiles $B_k$. The new category, $C'_k$, is delimited by taking into account the lower set $B_{k-1}$ and the upper set $B_{k+1}$. The new boundaries are characterized in the following way: $B'_h = B_h$ for $h = 0, \ldots, k-1$, $B'_{h-1} = B_h$ for $h = k+1, \ldots, M$.

b) Splitting: A splitting operation makes a separation of a category $C_k$ into two new adjacent categories, $C'_k$ and $C'_{k+1}$. This operation consists of adding a new boundary $B'_k$. The elements of $B'_k$ must fulfill Conditions 2 and 3. The new sets are characterized as follows: $B'_h = B_h$ for $h = 0, \ldots, k-1$, $B'_k = B'_k$, and $B'_h = B_{h-1}$ for $h = k+1, \ldots, M+1$.

**Definition 5 (Stability property)**

A method is considered stable under the operations of Definition 4, if and only if:

i) After performing a merging or a splitting operation, the actions belonging to a non-modified category previously to the change will keep their assignments after such a

---

[2] We keep the adjective "structural" used in (Almeida-Dias et al., 2010) and (Fernández et al., 2017). We make use also of the term "consistency properties".

modification.

ii) After performing a merging of two categories, the actions belonging to the merged categories (before merging) are still belonging to the new category.

iii) After performing a splitting operation of a category, the actions belonging to the modified category (before splitting) are still belonging to one of the two new categories.

The $S$-based primal and dual procedures fulfill a set of consistency properties described by the following three propositions:

**Proposition 3 (Properties under basic requirements)**
Under the basic requirements from Condition 2, the $S$-based primal and dual procedures fulfill the following properties:

i. *Uniqueness*: Each action is assigned to a unique class.
ii. *Independence*: The assignment of an action does not depend on the assignment of the other actions.
iii. *Homogeneity*: Actions which compare the same way with respect to the limiting boundaries are assigned to the same class.
iv. *Monotonicity*: $yDx$, and $x$ is classified into $C_k$, $\Rightarrow y$ is classified into $C_{k'}$ with $k' \geq k$.

(See the proof in Appendix A).

**Remark 2:**
If $S$ is transitive, $ySx$, and $x$ belongs to class $C_k$, $\Rightarrow y$ belongs to $C_{k'}$ with $k' \geq k$. The proof is similar to the one for Monotonicity, replacing $D$ by $S$ (see Appendix A).

**Proposition 4 (Conformity)**
Under Condition 2 and the $S$-based separability conditions (Conditions 3.i-3.iv), the $S$-based primal and dual procedures fulfill:
a) If $x$ belongs to $B_{Lk'}$, $x$ is assigned to $C_{k'+1}$;
b) If $x$ belongs to $B_{Uk'}$, $x$ is assigned to $C_{k'}$.

(See the proof in Appendix A).

**Proposition 5 (Stability Property)**

Under Condition 2 and the *D*-based separability requirements (Conditions 3.v-viii), the *S*-based primal and dual procedures are stable under merging and splitting operations.

(See the proof in Appendix A).

## 3. A *P*-based primal and dual procedures

In several papers, Bouyssou and Pirlot (2013, 2015a,b) have proved that the properties of the *P* relation used by the pseudo-disjunctive ELECTRE TRI are significantly different from the ones of the outranking relation *S* in the ELECTRE methods and more complex to analyze (Bouyssou and Marchant, 2015). Nevertheless, these conclusions are not necessarily valid in the general context of Condition 1, where *S* may be a relation different from the outranking relation in ELECTRE methods. In our view, the *P*-based primal and dual procedures, equivalent through the transposition operation, could be an alternative to the *S*-based conjoint procedure. To make it clearer our point, let us consider only the case of two ordered classes $C_1$ and $C_2$ and suppose that *b* is the single limiting action. *xSb* justifies assigning *x* to $C_2$ for the *S*-based primal procedure. But having *xSb* and *not(bSx)* ⇔ *xPb* reinforces the arguments in favour of assigning *x* to $C_2$. Under this view, a *P*-based assignment procedure can bring more information than the *S*-based, what may imply more justified assignments.

**3.1 The method**

Let us assume the same basic requirements imposed on the set of limiting profiles in Condition 2. In addition, we require the separability requirements stated by Condition 4.

**Condition 4 (Separability conditions)**

   i) There is no pair $(w,z) \in B_{Uk}$ $B_{Lk}$ fulfilling *wSz*;
   ii) There is no pair $(w,z) \in B_k$ $B_h$ (*h>k*) fulfilling *wSz*;
   iii) For each $z \in B_{Uk}$ there is $y \in B_{Lk}$ such that *ySz*;
   iv) For each $z \in B_k$ there is $w \in B_{k+1}$ such that *wDz*;
   v) For each $z \in B_k$ there is $w \in B_{k-1}$ such that *zDw*

**Remark 3:**

Conditions 4.ii, 4.iv and 4.v are equivalent to the separability conditions in ELECTRE TRI-nB (cf. Fernández et al., 2017). From Conditions 4.i and 4.iii, it follows that each action *z* in $B_{Uk}$ fulfills *yPz* with some action $y \in B_{Lk}$, what is a preference-based separability condition between actions within the same limiting boundary $B_k$.

**Definition 6 (*P* relation between actions and boundaries)**

a) $xPB_k \Leftrightarrow$ There is $w \in B_k$ such that $xPw$ and there is no $z \in B_k$ fulfilling $zPx$;

b) $B_k Px \Leftrightarrow$ There is $w \in B_k$ such that $wPx$ and there is no $z \in B_k$ fulfilling $xPz$

**Definition 7 (*P*-based primal assignment procedure)**

Set the relational system $(D,P)$ according to Condition 1. Set $B_k$, $k=1,…M$-1, fulfilling Condition 2 and take $B_M$ as the ideal action. Set $B_M Px$ for all $x$.

- For $k=1,…M$, find the first limiting boundary $B_k$ fulfilling $B_k Px$;
- Take $C_k$ as an acceptable category to assign $x$.

**Definition 8 (*P*-based dual assignment procedure)**

Set the relational system $(D,P)$ according to Condition 1. Set $B_k$, $k=1,…M$-1, fulfilling Condition 2 and take $B_0$ as the anti-ideal action. Set $xPB_0$ for all $x$.

- For $k=M$-1,…,0, find the first $B_k$ fulfilling $xPB_k$;
- Take $C_{k+1}$ as an acceptable category to assign $x$.

**Remark 4:**

Both rules can work if $B_{Uk}$ (respectively $B_{Lk}$) is empty for some $k$, even for all $k$ from 1 to $M$-1. The importance of having non-empty $B_{Uk}$ and $B_{Lk}$ is related to the Conformity Property (see Proposition 8 and its proof). If $B_{Uk}$ is empty for all $k$, the primal assignment rule is basically identical to the ELECTRE TRI-nB and INTERCLASS-nB pseudo-disjunctive procedure.

**Proposition 6 (Basic properties of the *P* relation between actions and limiting boundaries)**

Under the *D*-based separability requirements (Condition 4.iv and 4.v), we have:
i) $xPB_k \Rightarrow xPB_h$ for $k>h$;
ii) $B_k Px \Rightarrow B_h Px$ for $h>k$

(See the proof in Appendix A).

## 3.2 Structural properties of the *P*-based primal and dual procedures

As in Subsection 2.3, we should prove the fulfillment of the consistency properties suggested by Roy and Bouyssou (1993).

**Proposition 7 (Properties under basic requirements)**

Under the basic requirements from Condition 2, the *P*-based primal and dual procedures fulfill the following properties:

- *Uniqueness*: Each action is assigned to a unique class.
- *Independence*: The assignment of an action does not depend on the assignment of the other actions.
- *Homogeneity*: Actions which compare the same way with respect to the limiting boundaries are assigned to the same class.
- *Monotonicity: yDx*, and *x* is assigned to $C_k$, ⇒ *y* is assigned to $C_{k'}$ with $k' \geq k$.

(See the proof in Appendix A).

**Proposition 8 (Conformity)**

Under Conditions 2 and 4, the *P*-based primal and dual procedures fulfill:

- If *x* belongs to $B_{Lk'}$, *x* is classified into $C_{k'+1}$;
- If *x* belongs to $B_{Uk'}$, *x* is assigned to $C_{k'}$

*Proof:*

Primal rule:

Suppose that $x \in B_{Lk'}$. From Conditions 2.i and 4.i, there is no *w* in $B_{k'}$ such that *wPx*. Hence, *not*($B_{k'}Px$) is fulfilled. From Conditions 4.ii and 4.iv, there is *z* in $B_{k'+1}$ fulfilling *zPx* and there is no $w \in B_{k'+1}$ such that *wPx*. Hence, $B_{k'+1}Px$ (Definition 7). This implies that *x* is assigned to $C_{k'+1}$ by the primal procedure.

Suppose now that $x \in B_{Uk'}$. From Condition 4.i, there is no *w* in $B_{k'-1}$ fulfilling *wPx*. It follows that *not*($B_{k'-1}Px$). From Condition 4.iii there is *w* in $B_{k'L}$ such that *wPx*; additionally, from Conditions 2.iv and 4.i there is no *z* in $B_{k'}$ such that *xPz*. Hence, $B_{k'}Px$ and *x* is classified into $C_{k'}$ by the primal rule.

The proof for the dual rule is omitted. It can be justified by the equivalence through the transposition operation.

**Remark 5:**

If $B_{Lk'}$ is empty, Condition 4.iii cannot be applied and the property is not fulfilled by $x \in B_{Uk'}$. Similarly, for the dual assignment rule, if $B_{Uk'}$ is empty, the lack of Condition 4.iii impedes to prove that $x \in B_{Lk'} \Rightarrow xPB_{k'}$, so $x$ is not necessarily assigned to $C_{k'+1}$.

**Proposition 9 (Stability)**

Under Condition 2 and the $D$-based separability requirements (Conditions 4.iv and 4.v), the $P$-based primal and dual procedures are stable under merging and splitting operations.

The proof follows a similar logic as in Proposition 5, replacing Definitions 2 and 3, and Proposition 2.i and 2.ii by Definitions 7 and 8, and Propositions 6.i and 6.ii.

## 4. Two illustrative examples

Let us start with a toy example. Consider a set of actions $A = [0, 3]^5$ (preference increasing with criterion values), and three ordered classes. Suppose that the pair DM-decision analyst elicits the parameters of an ELECTRE TRI model as follows below:

- Equal criterion weights $w_i = 0.2$, $i=1,\ldots5$;
- Indifference thresholds $q_i = 0$, $i=1,\ldots5$;
- Preference thresholds $p_i = 0.5$, $i=1,\ldots5$;
- Veto thresholds $v_i = 1.5$, $i=1,\ldots5$;
- Pre-veto thresholds (as in Mousseau and Dias, 2004) $u_i = 1$, $i=1,\ldots5$;

The DM should set two limiting boundaries $B_1$ and $B_2$. The actions in $B_1$ and $B_2$ are denoted $b_{Uk,j}$, $b_{Lk,j}$, $k=1,2$. The performances of the limiting actions are shown in Table 1.

**Table 1. Performances of limiting actions**

|   | $g_1$ | $g_2$ | $g_3$ | $g_4$ | $g_5$ |
|---|---|---|---|---|---|
|   | 1 | 1 | 1 | 1 | 0.5 |
| $b^U_{1,1}$ | | | | | |
| $b^L_{1,1}$ | 1 | 1 | 2.5 | 0.5 | 0.5 |
| $b^U_{2,1}$ | 1 | 1 | 2 | 2 | 1 |
| $b^L_{2,1}$ | 2.5 | 1 | 2.5 | 1.5 | 0.5 |
| $b_{L2,2}$ | 1 | 2.5 | 2.5 | 1.5 | 0.5 |

Let us take the dominance relation as $D$. Set $\lambda = 0.6$ as credibility threshold to establish the crisp outranking relation. $S$, $D$, and $P$ relations between limiting actions are given in Table 2.

**Table 2. Binary relations between limiting actions**

| | $b_{L2,1}$ | $b_{L2,2}$ | $b_{U2,1}$ | $b_{L1,1}$ | $b^U_{1,1}$ |
|---|---|---|---|---|---|
| $b_{L2,1}$ | $S$ | $Inc$ | $P$ | $D, P$ | $D, P$ |
| $b_{L2,2}$ | $Inc$ | $S$ | $P$ | $D, P$ | $D, P$ |
| $b_{U2,1}$ | $P^{-1}$ | $P^{-1}$ | $S$ | $P$ | $D, P$ |
| $b_{L1,1}$ | $D^{-1}, P^{-1}$ | $D^{-1}, P^{-1}$ | $P^{-1}$ | $S$ | $P$ |
| $b_{U1,1}$ | $D^{-1}, P^{-1}$ | $D^{-1}, P^{-1}$ | $D^{-1}, P^{-1}$ | $P^{-1}$ | $S$ |

**Note:**

$xR^{-1}y \Leftrightarrow yRx$;

$xIncy \Leftrightarrow not(xSy)$ and $not(ySx)$

In Table 2, it is easy to check that the limiting actions in Table 1 fulfill Condition 3 and Condition 4. Hence, both the $S$-based and the $P$-based conjoint procedure can be applied, also fulfilling the structural properties given by Propositions 3-5 and 7-9.

Let $x = (2, 1, 2, 1, 2)$ be an action to be assigned. With the limiting profiles, $x$ fulfills the relations shown in Table 3.

**Table 3. Binary relations between the action and limiting profiles**

|   | $b_{U\,1,1}$ | $b_{L\,1,1}$ | $b_{U2,1}$ | $b_{L\,2,1}$ | $b_{L\,2,2}$ |
|---|---|---|---|---|---|
| $x$ | $P$ | $P$ | $S/S^{-1}$ | $Inc$ | $Inc$ |

The relations between the action and the limiting boundaries (Definitions 1 and 6) are provided by Table 4.

**Table 4. Relations between the action and the limiting boundaries**

|   | $B_0$ | $B_1$ | $B_2$ | $B_3$ |
|---|---|---|---|---|
| $x$ | $S$ | $S$ | $S^{-1}$ | $S^{-1}$ |
| $x$ | $P$ | $P$ | $Inc_P$ | $P^{-1}$ |

**Note:**

$xR^{-1}B \Leftrightarrow BRx;$

$xInc_PB \Leftrightarrow not(xPB)$ and $not(BPx)$

According to Definitions 2 and 3, $x$ is assigned to $C_2$ by both the $S$-based primal and dual rules. This is consistent with the fact that $x$ is indifferent to $b_{U2,1}$, which belongs to the upper "layer" of $C_2$. From Definition 7, $x$ is assigned to $C_3$ by the $P$-based primal rule, and to $C_2$ by $P$-based dual rule (Definition 8). Thus, the $P$-based conjoint procedure suggests $C_2$ and $C_3$ as possible assignments for $x$. With the information provided by $P$, the method is unable to suggest a well-defined assignment. From its indifference with $b_{U2,1}$, it follows that $x$ is preferentially close to the limiting boundary $B_2$; since also $not(B_2Px)$, the hesitation between $C_2$ and $C_3$ is justified.

Let us address below a realistic size example.

The integrated impact of a set of Research and Development projects is determined by four points of view: $g_1$= "social impact", $g_2$= "improvement of the research team competence", $g_3$= "economic impact", and $g_4$= "scientific impact". The criterion performances range in $[0, 8]$. The aggregated impact is evaluated in the set of ordered classes $C=$ {Very Low ($C_1$), Low ($C_2$), Below Average ($C_3$), Average ($C_4$), Above Average ($C_5$), High ($C_6$), Very High ($C_7$), Outstanding ($C_8$)}.

The DM sets an ELECTRE model with the following parameters:

Indifference thresholds: $q_1=0.1, q_2=0.3, q_3=0.2, q_4=0.1$;

Preference thresholds: $p_1=1.2, p_2=1.7, p_3=1.8, p_4=1.1$;

Pre-veto thresholds: $u_1=2.1, u_2=2.6, u_3=2.7, u_4=2.1$;

Veto thresholds: $v_1=2.5, v_2=3.1, v_3=3.1, v_4=2.9$;

Weights: $w_1=0.24, w_2=0.23, w_3=0.27, w_4=0.26$;

Credibility threshold $\lambda=0.85$

The DM should set seven limiting boundaries. The performances of the limiting actions (two per boundary) are shown in Table 5.

**Table 5. Performances of limiting actions (second example)**

|  | $g_1$ | $g_2$ | $g_3$ | $g_4$ | *Overall Impact* |
|---|---|---|---|---|---|
| $b^U_{1,1}$ | 0.5 | 2 | 1 | 0.5 | Very Low |
| $b^L_{1,1}$ | 1 | 0.5 | 0 | 1 | Low |
| $b^U_{2,1}$ | 2.5 | 2 | 1 | 1.5 | Low |
| $b^L_{2,1}$ | 1 | 1 | 2 | 2 | Below Average |
| $b^U_{3,1}$ | 2.5 | 2.5 | 2 | 2 | Below Average |
| $b^L_{3,1}$ | 2.5 | 2 | 3.5 | 2.5 | Average |
| $b^U_{4,1}$ | 4.5 | 3.5 | 5.5 | 3.5 | Average |
| $b^L_{4,1}$ | 5.5 | 3 | 4 | 3 | Above Average |
| $b^U_{5,1}$ | 6 | 6 | 6 | 4 | Above Average |
| $b^L_{5,1}$ | 7 | 4 | 5.5 | 3.5 | High |
| $b^U_{6,1}$ | 6.5 | 7 | 6.5 | 4 | High |
| $b_{L6,1}$ | 7 | 5.5 | 6 | 4.5 | Very High |
| $b_{U7,1}$ | 7 | 7.5 | 7 | 5.5 | Very High |
| $b_{L7,1}$ | 8 | 7.5 | 6.5 | 7 | Outstanding |

Let us take the Pareto dominance as $D$. With the above ELECTRE model's parameters, the limiting actions in Table 5 satisfy Conditions 2 and 3. Condition 2 is obviously fulfilled. The $D$-based separability conditions can be easily checked in Table 5. We will focus on Condition 3.i. To check this, we provide below the values of $\sigma(b_{Uk,1}, b_{Lk,1})$, $k=1,\ldots 7$.

$\sigma(b_{U1,1}, b_{L1,1}) = 0.809$, $\sigma(b_{U2,1}, b_{L2,1}) = 0.761$, $\sigma(b_{U3,1}, b_{L3,1}) = 0.677$, $\sigma(b_{U4,1}, b_{L4,1}) = 0.804$, $\sigma(b_{U5,1}, b_{L5,1}) = 0.804$, $\sigma(b_{U6,1}, b_{L6,1}) = 0.809$, $\sigma(b_{U7,1}, b_{L7,1}) = 0.544$;

With $\lambda = 0.85$ we have $not(b_{Uk,1}Sb_{Lk,1})$ for $k=1,\ldots 7$.

Let $x = (4, 4, 4, 4)$ be an action to be assigned. In order to apply the $S$-based primal rule, in a descending procedure from $k=7$ we should identify the first $k$ for which $xSb_{Lk,1}$. We have:

$\sigma(x, b_{Lk,1}) = 0$ for $k=7, 6, 5$; $\sigma(x, b_{L4,1}) = 0.76 < \lambda$; $\sigma(x, b_{L3,1}) = 1$;

It is obvious that $\sigma(x, b_{U3,1}) = 1 \Rightarrow not(b_{U3,1}Px)$; hence, we have both $xSB_3$ and $not(xSB_4)$ (Definition 1). From Definition 2, $x$ is assigned to $C_4$ ("Average").

Now, applying the dual rule, in an ascending procedure from $k=1$ we should find the first $k$ such that $b_{Uk,1}Sx$; we have:

$\sigma(b_{Uk,1}, x)=0$, $k=1, 2$; $\sigma(b_{U3,1}, x)=0.033$; $\sigma(b_{U4,1}, x)= 0.863 > \lambda$; additionally $\sigma(x, b_{L4,1}) = 0.76 < \lambda$; hence, from Definition 1 it follows that $B_4Sx$ and $not(B_3Sx) \Rightarrow x$ is assigned to $C_4$ by the dual rule.

## 5. Discussion

We can question whether the requirements of the methods are too restrictive. Condition 1 establishes requirements to the model of preferences, whereas Conditions 2, 3, and 4 make requirements to the decision maker. Condition 2 establishes minimal requirements, which are not different from those of ELECTRE TRI-nB, except because we should use two "layers" to describe each limiting boundary. Requirements in Condition 3 (for the $S$-based conjoint procedure) may be questioned because its number, although 3.ii, 3.iii, and 3.iv arise naturally from the order of classes. Requirements in Condition 4 are similar to the hyper-separability condition in ELECTRE TRI-nB (see Fernández *et al.*, 2017), except for the preference-based separability requirement underlined by Remark 3, which demands a high discrimination power from the decision maker.

As the numbers of classes and limiting actions increase, a direct elicitation of these actions with the fulfillment of all the requirements becomes more difficult. Monotonicity, homogeneity, independence and uniqueness, perhaps the most important structural properties, are fulfilled under very basic conditions. Other requirements could be weakened. Without some of the requirements established by Conditions 1-3 (respectively Condition 4), the *S*-based (resp. *P*-based) primal-dual conjoint procedure still works, but fails to satisfy certain structural properties. Let us analyze some issues:

- If *S* is not reflexive, Conformity is not fulfilled by the *S*-based procedure. This is the case when, using the interval outranking approach, the criterion performances of the limiting actions are non-degenerate interval numbers;
- Conditions 3.i-3.iv (respectively, 4.i-4.iii) are introduced to guarantee Conformity in the *S*-based (resp. *P*-based) procedure. If any of these requirements is not fulfilled, Conformity is not guaranteed for the entire set of limiting actions. Nevertheless, the assignment method still works for many other actions of the universe. For instance, if there is a pair $(w,z) \in B_{Uk} \; B_{Lk}$ fulfilling *wSz* (in contradiction with Condition 3.i), Conformity is not fulfilled for action *w*, but is kept for the other limiting actions; from a practical point of view, this non-fulfillment is important only if some actions to be assigned are similar to *w*.
- Without the *D*-based separability conditions, Stability is not fulfilled by the *S*-based procedure; from a practical point of view, the non-fulfillment of this property is only important if the DM hesitates about the most appropriate definition of the set of classes; once this set has been convincingly defined, Stability becomes a theoretical curiosity. However, the *D*-based separability conditions are required to prove Propositions 2 and 6. Without Proposition 2, using the *S*-based primal rule we could have $xSB_k$ and $not(xSB_h)$ for some $h < k$. This questions assigning *x* to a class not worse than $C_k$, as suggested by the rule. Something similar happens to the *S*-based dual rule and the *P*-based conjoint procedure.
- Without the *D*-based separability conditions, Stability and Conformity are not satisfied by the *P*-based procedure;
- The *P*-based procedure still works when some upper or lower limiting boundaries (one of them for each boundary) is empty, although Conformity fails to be fulfilled. The *S*-based procedure requires both non-empty subsets $B_{Uk}$ and $B_{Lk}$.

For each relational system of preferences fulfilling Condition 1, if the decision maker is willing to fulfill Conditions 2 and 3 (respectively 2 and 4), the pair DM-decision analyst can use Definitions 1-3

(resp. 6-8) to build an outranking-based (resp. *P*-based) primal-dual conjoint procedure compatible with Propositions 3-5 (resp. 7-9). This remark confers a wide generality to our proposal, beyond ELECTRE-type methods, even beyond multi-criteria decision methods. We will distinguish some important particular cases:

1. Classical ELECTRE framework

    This case arises when *S* is the a crisp outranking relation obtained from the credibility index of the outranking $\sigma(x,y)$ used by the later ELECTRE methods (Roy, 1991). Let $\delta$ denote a real number within $]0.5, 1]$ considered as a credibility threshold to establish the crisp preference relations. $xSy \Leftrightarrow \sigma(x,y) \geq \delta$; *P* is defined by Condition 1[3] and *D* is the classical Pareto dominance relation. It is well-known that such a relational system fulfills Condition 1. Hence, under Condition 2 the *S*-based (respectively, the *P*-based) conjoint primal-dual procedure given by Definitions 2-3 (resp. 7-8) can work, and under Condition 3 (resp. 4) all the structural properties of Section 2 (resp. 3) given by Propositions 3-5 (resp. 7-9) are fulfilled. If in Condition 2, we set $B_{Uk}$ as empty for $k=1,...M-1$, it is easy to prove that the *S*-based (resp. *P*-based) primal procedure is equivalent to the pseudo-conjunctive (resp. pseudo-disjunctive) method of ELECTRE TRI-nB, and ELECTRE TRI-B if $card(B_k)=1$ for $k=1,...M-1$.

2. Non-compensatory sorting models

    This case is similar to the above. Under the axiomatic bases studied by Bouyssou and Marchant (2007), a majority sorting rule can be used to assign actions to ordered classes. Instead of the credibility index of the outranking, a majority index value allows to establish a crisp outranking relation *S* (e.g. Meyer and Olteanu, 2019). Combined with Pareto dominance, *S* fulfills Condition 1.

3. ELECTRE framework with interacting criteria

    ELECTRE methods originally required a family of criteria to be defined where no interaction between any pair of criteria exists. This could be a real limitation since the interaction between criteria naturally arises in many situations. Contemplating these situations, Figueira et al. (2009) presented a rather straightforward way to adapt the concordance index, a crucial

---

[3] In all the cases, *P* is defined from *S* and Condition 1.

component of ELECTRE methods, in such a way that a significant interaction between any pair of criteria can be considered if it exists and ignored if it does not exist. In the approach by Figueira et al. (2009), the concordance index increases when there are pairs of criteria with strengthening interaction in the concordance coalition, and decreases when there are pairs of criteria with weakening interaction in the same coalition. It also decreases when there is antagonism between criteria in concordance coalition and criteria in discordance coalition. Let $(x,y)$ denote a pair of actions belonging to $A \times A$. Let $G$ denote the set of criteria; $g_i(x)$ denotes the evaluation of action $x$ on criterion $g_i$. We suppose preference increasing with criterion values. Figueira et al. (2009) proved that the extended concordance index $c(x, y)$ is a non-decreasing function of $g_i(x) - g_i(y)$ for all criterion $g_i$. As in other ELECTRE methods, the credibility index of the outranking is calculated as $\sigma_l(x,y) = c(x, y) \cdot (1-d(x,y))$ (where $d(x,y)$ is the discordance index). $d(x,y)$ is obtained from a certain aggregation of the marginal discordance indexes. This aggregation should fulfill that, for all $g_i$, $d(x,y)$ is non-increasing with respect to $g_i(x)-g_i(y)$. A crisp outranking relation is defined as $xSy \Leftrightarrow \sigma_l(x,y) \geq \delta$. Let us take $D$ as the Pareto dominance. The pair $(D,S)$ fulfills Condition 1 as proved below:

i. $xDy \Rightarrow xSy$

$xDy \Rightarrow g_i(x) \geq g_i(y) \; \forall \; g_i \in G \Rightarrow c(x,y)=1$ and $d(x,y)=0 \Rightarrow \sigma_l(x,y)=1 \Rightarrow xSy$

ii. $xSy$ and $yDz \Rightarrow xSz$

$yDz \Rightarrow g_i(y) \geq g_i(z) \; \forall \; g_i \in G \Rightarrow -g_i(y) \leq -g_i(z) \; \forall \; g_i \in G \Rightarrow g_i(x)-g_i(y) \leq g_i(x)-g_i(z) \; \forall \; g_i \in G$. Since $c(x,y)$ is non-decreasing with respect to $g_i(x)-g_i(y)$ then

$c(x,z) \geq c(x,y)$ ……………………………………………………………………… (a)

Since $g_i(x)-g_i(y) \leq g_i(x)-g_i(z) \; \forall \; g_i \in G$ and $d(x,y)$ is non-increasing with respect to $g_i(x)-g_i(y)$, then

$d(x,z) \leq d(x,y)$ ……………………………………………………………………. (b)

From (a), (b) and the definition of $\sigma_l(x,y)$ we have that $\sigma_l(x,y) \leq \sigma_l(x,z) \Rightarrow xSz$.

iii. $xDy$ and $ySz \Rightarrow xSz$

$xDy \Rightarrow g_i(x) \geq g_i(y) \; \forall \; g_i \in G \Rightarrow g_i(x)-g_i(z) \geq g_i(y)-g_i(z) \; \forall \; g_i \in G$. Since $c(x,y)$ is non-decreasing with respect to $g_i(x)-g_i(y)$ we have

$c(x,z) \geq c(y,z)$ ……………………………………………………………………... (c)

Since $g_i(x)-g_i(z) \geq g_i(y)-g_i(z)\ \forall\ g_i \in G$ and $d(x,y)$ is non-increasing with respect to $g_i(x)-g_i(y)$ then

$$d(x,z) \leq d(y,z) \quad \text{...............................................................} \quad (d)$$

From (c), (d) and the definition of $\sigma_l(x,y)$, it follows that $\sigma_l(y,z) \leq \sigma_l(x,z) \Rightarrow xSz$.

This conclusion allows to handle interacting criteria in both ELECTRE TRI-B and ELECTRE TRI-nB, without losing their structural properties.

4. Hierarchical ELECTRE

In handling hierarchical structures under the ELECTRE paradigm, a crisp outranking relation $S_h$ can be set on non-elementary criteria $g_h$ as in Corrente et al. (2013, 2016, 2017). Combining this relation with a preference relation $xP_hy \Leftrightarrow xS_hy$ and $not(yS_hx)$, and the Pareto dominance on subsets of elementary criteria which are descending from $g_h$, under the same arguments given above we can have a relational system of preferences which fulfils Condition 1. Thus, defining appropriately the limiting boundaries between classes according to Condition 2, the $S$-based (resp. $P$-based) primal-dual assignment procedure can work on the highest hierarchical level, or on a chosen non elementary sub-criterion. Under Condition 3 (resp. 4), the assignment procedure fulfills the consistency properties from Propositions 3-5 (resp. 7-9). The hierarchical pseudo-conjunctive and pseudo-disjunctive ELECTRE TRI-B with interacting criteria proposed by Corrente et al. (2016) are respectively particular cases of the $S$-based and $P$-based primal procedures when $B_{Uk}$ is empty and card($B_{Lk}$) =1 for $k=1,…M-1$.

5. Interval outranking approach

This is an extension of the outranking approach to the interval framework proposed by Fernández et al. (2019, 2020). Imprecisions on weights and veto thresholds are handled by using interval numbers. Imprecision and uncertainty on criterion performance levels can be handled by interval numbers in some criteria and discriminating thresholds for other criteria. For a given majority threshold (expressed by an interval number), the method obtains a degree of credibility of the interval outranking. Let $L(x,y,\lambda)$ denote the credibility index of the interval outranking with a majority threshold $\lambda$. A crisp outranking relation can be defined as $xSy \Leftrightarrow L(x,y,\lambda) \geq \delta$ Let us denote by $G_1$ the set of criteria whose levels are represented by pseudo-criteria; $G_2$ denotes the set of criteria whose scores are represented by non-degenerate interval numbers. Assume preference increasing with criterion values. The interval

dominance is defined as follows:

**Definition 9 (Interval Dominance)**

Let $(x,y) \in A \times A$, where $x \neq y$, and $G_2$ is not empty. $y$ is $\alpha$-dominated by $x$ iff the following is true:

i)      $g_j(x) \geq g_j(y)$, for all $g_j \in G_1$,

ii) $min\{Poss(\boldsymbol{g_j}(x) \geq \boldsymbol{g_j}(y))$, for all $g_j \in G_2\} \geq \alpha \geq 0.5$

where the italic bold letter denotes interval numbers, and *Poss* is a possibility function given by:

$$\quad (1)$$

$\boldsymbol{B} = [b^-, b^+]$ and $\boldsymbol{C} = [c^-, c^+]$ are interval numbers and $p_{BC} = \dfrac{b^+ - c^-}{(b^+ - b^-) + (c^+ - c^-)}$.

In (Fernández et al., 2019, 2020), the above possibility function is interpreted as a degree of credibility of $\boldsymbol{B} \geq \boldsymbol{C}$.

If $G_2$ is empty, under Condition i) above, it is said that $y$ is 1-dominated by $x$. The 1-dominance is identical to the classical notion of Pareto dominance.

Fernández et al. (2019) proved that the interval dominance and the interval outranking satisfy the requirements of Condition 1.i-1.iii. Although $S$ is not reflexive on interval numbers, if $x$ is described by pseudo-criteria or by degenerate interval numbers, $xSx$ is fulfilled, what permits to satisfy the Conformity Property of the $S$-based method when the criterion performances of the limiting actions in Conditions 2.i and 2.ii are real numbers.

INTERCLASS-nB, a generalization of ELECTRE TRI-nB using the interval outranking, was proposed in (Fernández et al., 2020). Again if in Condition 2 we set $B_{Uk}$ as empty for $k=1,\ldots M-1$, it is easy to prove that the $S$-based (respectively, $P$-based) primal procedure is equivalent to the pseudo-conjunctive (respectively, pseudo-disjunctive) rule of INTERCLASS-nB.

6. Hierarchical interval outranking with interacting criteria

The interval outranking proposed by Fernández et al. (2019) may be extended to handle interacting criteria and hierarchical structures. Suppose that $S_h$ is a crisp outranking relation

7. In a PROMETHEE framework

    Let $\Pi(x,y)$ denote the binary preference degree calculated by PROMETHEE. A reflexive outranking relation can be defined as $xSy \Leftrightarrow \Pi(x,y) - \Pi(y,x) \geq 0$. If $D$ is the Pareto dominance relation, it is easy to prove that the relational system $(D, S)$ satisfies Condition 1.

8. Imprecise value functions

    Suppose that the decision maker's preferences are modeled by an interval valued function $U$; its simplest case is an interval weighted sum function, in which criterion weights and criterion scores are interval numbers. A reflexive outranking relation can be defined as $xSy \Leftrightarrow Poss\ (U(x) \geq U(y)) \geq 0.5$, where $Poss$ is the possibility function from Equation 1. $S$ is reflexive. A stronger $D$ relation can be defined as $xDy \Leftrightarrow Poss\ (U(x) \geq U(y)) \geq \alpha > 0.5$, fulfilling Condition 1.

9. Partial compensatory multiple criteria decision models

    Let us consider a multiple criteria decision model in which an ordinal value function $U$ is complemented by veto conditions in some or all criteria. That is, for all pair of actions $(x,y)$, $U(x) \geq U(y)$ and no veto condition is fulfilled $\Leftrightarrow$ x is at least as good as y $\Leftrightarrow$ $xSy$ $S$ is reflexive. Again taking $D$ as the Pareto dominance relation, the pair $(D,S)$ fulfills Condition 1.

10. In a group decision framework

    If the property $\Xi$ is a certain measure of consensus, with an appropriate definition of $(D,S)$, the $S$-based (resp. $P$-based) primal- dual conjoint procedure could be used to assign potential collective decisions to classes of acceptable agreement.

## 6. Conclusions

This paper has presented a general approach to designing ordinal classification methods based on comparing actions with limiting boundaries of ordered classes. Each boundary is described by two subsets of limiting actions, making each class closed below and above. This avoids the contradiction observed by other authors between the equivalence through the transposition operation and the assignment of limiting actions to the classes to which they belong, closing the discussion about what property should predominate. The methods require a relational system $(D,S)$, where $S$ is a reflexive relation, compatible with the order of classes, and *D* is a transitive relation stronger than *S*. From *S*, an asymmetric preference relation *P* is defined. On this background, we propose *S*-based and *P*-based assignment methods. Each is composed of two complementary assignment procedures, which correspond through the transposition operation and should be used conjointly. The methods work under several basic conditions on the set of limiting boundaries. Under other more demanding separability requirements, each procedure fulfills the structural properties of Conformity, Stability, Monotonicity, Homogeneity, Independence and Uniqueness. Nevertheless, the methods still work without some of the demanding conditions, although losing some properties. The pseudo-conjunctive (respectively, pseudo-disjunctive) assignment methods of ELECTRE TRI-B, ELECTRE TRI-nB, INTERCLASS-nB, and the hierarchical ELECTRE TRI-B with interacting criteria, are particular cases of the *S*-based (resp. *P*-based) descending (resp. ascending) assignment rule proposed here. Using this general approach, *S* and *P*-based ordinal classification procedures with desirable properties arise from each decision method with capacity to build preference relations as in Condition 1. This gives a theoretical support to the introduction of many diverse methods based on limiting boundaries between adjacent classes. This point was illustrated by many different kinds of decision models (Section 5); as an avenue of future research, some of them could be subject of study in forthcoming papers.

The *S* and *P* based procedures are alternative. The *P*-based method uses more information to suggest assignments, requires a smaller number of separability conditions, and can work with classes opened below or above, although losing the Conformity Property; however, it requires a strong preference separability condition between actions in the same boundary belonging to adjacent classes. The selection of the most appropriate method may be depending on i) the way in which *S* is defined; ii) the nature of criterion scales and cardinality of the decision set; iii) the method (direct or indirect) used to elicit the model's parameters, including the limiting actions; and iv) the specific characteristics of the decision maker. Finding guidelines to determine the most appropriate method is a second avenue for future research.

An illustrative and simple example showed that a direct setting of a few limiting actions fulfilling all the requirements could not be a demanding cognitive task for the decision maker. A real size example with eight classes, although addressed satisfactorily, required much more effort. As the number of classes increases, setting many limiting actions fulfilling all the conditions would require indirect elicitation methods, what is a third direction of research.

**Acknowledgments**

Eduardo Fernández (respectively, Jorge Navarro) is grateful for the support from the Autonomous University of Coahuila (resp. Sinaloa). José Rui Figueira thanks the support by Fundação para a Ciência e a Tecnologia (FCT) in the framework of the project PTDC/EGE-OGE/30546/2017 (hSNS: Portuguese public hospital performance assessment using a multi-criteria decision analysis framework).

**References**

Almeida-Dias, J., Figueira, J., and Roy, B. (2010). ELECTRE TRI-C: A multiple criteria sorting method based on characteristic reference actions. *European Journal of Operational Research,* 204: 565-580.

Almeida-Dias, J., Figueira, J. R., and Roy, B. (2012). A multiple criteria sorting method where each category is characterized by several reference actions: The ELECTRE TRI-nC method. *European Journal of Operational Research*, 217(3): 567-579.

Araz, C., and Ozkarahan, I. (2007). Supplier evaluation and management system for strategic sourcing based on a new multicriteria sorting procedure. *International Journal of Production Economics*, 106: 585-606.

Bouyssou, D. and Marchant, T. (2007). An axiomatic approach to noncompensatory sorting methods in MCDM, I: The case of two categories. European Journal of Operational Research, 178(1):217-245.

Bouyssou, D., and Marchant, T. (2015). On the relations between ELECTRE TRI-B and ELECTRE TRI-C and on a new variant of ELECTRE TRI-B. *European Journal of Operational Research*, 242(1): 201-211.

Bouyssou, D. and Pirlot, M. (2013). On the relationship between strict and non-strict outranking relations. Cahier du LAMSADE 346, LAMSADE, Université Paris Dauphine. Available in http://lamsade.dauphine.fr/~bouyssou

Bouyssou, D. and Pirlot, M. (2015a). A consolidated approach to the axiomatization of outranking relations: A survey and new results. Annals of Operations Research, 229(1):159-212. doi: 10.1007/s10479-015-1803-y.

Bouyssou, D. and Pirlot, M. (2015b) A note on the asymmetric part of an outranking relation. International Transactions in Operational Research, 22(5):883-912. doi: 10.1111/itor.12135.

Bouyssou, D, Marchant, T., and Pirlot, M. (2020). A theoretical look at ELECTRE TRI-nB https://arxiv.org/abs/2008.09484

Corrente S, Greco S, Słowiński R. (2013). Multiple criteria hierarchy process with ELECTRE and PROMETHEE. *Omega: The International Journal of Management Science*, 41,820–46.

Corrente, S., Greco, S. and Słowiński, R. (2016). Multiple criteria hierarchy process for ELECTRE TRI methods. *European Journal of Operational Research*, 252(1): 191-203.

Corrente, S., Figueira, J. R., Greco, S. and Słowinski, R. (2017). A robust ranking method extending Electre III to hierarchy of interacting criteria, imprecise weights and stochastic analysis. *Omega-International Journal of Management Science*, 73, 1–17.

Fernandez, E., and Navarro, J. (2011). A new approach to multi-criteria sorting based on fuzzy outranking relations: The THESEUS method. *European Journal of Operational Research*, 213: 405-413.

Fernández, E., Figueira, J., Navarro, J. & Roy, B. (2017). ELECTRE TRI-nB: A new multiple criteria ordinal classification method. *European Journal of Operational Research*, 263 (1), 214-224.

Fernández, E., Figueira, J., and Navarro, J. (2019). An interval extension of the outranking approach and its application to multiple-criteria ordinal classification, *Omega: The International Journal of Management Science*, 84, 189-198.

Fernández, E., Figueira, J., & Navarro, J. (2020). Interval-based extensions of two outranking methods for multi-criteria ordinal classification, *Omega: The International Journal of Management Science*, 95, 102065 . https://doi.org/10.1016/j.omega.2019.05.001

Figueira, J.R., Greco, S., Roy, B. (2009). ELECTRE methods with interaction between criteria: an extension of the concordance index. *European Journal of Operational Research*,199(2), 478–95.

Govindan, K. and Jepsen, M. (2016). ELECTRE: A comprehensive literature review on methodologies and applications. *European Journal of Operational Research*,250(1):1-29.

Greco, S., Matarazzo, B., and Słowiński, R. (2001). Rough sets theory for multicriteria decision analysis. *European Journal of Operational Research,* 129: 1-47.

Ishizaka A, Nemery P, Pearman C. (2012). AHPSort: an AHP based method for sorting problems. International Journal of Production Research, 50: 4767-84.

Jacquet-Lagrèze, E. (1995): An application of the UTA discriminant model for the evaluation of R&D projects. In Pardalos, P.M. Siskos, Y., Zopounidis, C.(eds.) *Advances in Multicriteria Analysis*, Kluwer Academic Publishers, Dordrecht, The Netherlands. pp. 203-211.

Köksalan, M., and Ulu, C. (2003). An interactive approach for placing alternatives in preference categories. *European Journal of Operational Research,* 144: 429-439.

Massaglia, R., and Ostanello, A. (1991). N-TOMIC: A support system for multicriteria segmentation problems. In: Korhonen, P., Lewandowski, A., and Wallenius, J. (Eds.). *Multiple Criteria Decision Support*. LNEMS, Volume 356, Springer-Verlag, Berlin, Germany, pp. 167-174.

Meyer, P., and Olteanu, A. L. (2019). Handling imprecise and missing evaluations in multi-criteria majority-rule sorting. Computers & Operations Research, 110, 135-147.

Nemery, P., and Lamboray C. (2008), FlowSort: a flow-based sorting method with limiting or central profiles. *TOP*,16: 90-113.

Roy, B. (1991). The outranking approach and the foundations of ELECTRE methods. *Theory and Decision,* 31: 49-73.

Roy, B., and Bouyssou, D. (1993). *Aide Multicritère à la Décision : Méthodes et Cas*. Economica, Paris, France.

Roy, B. (2002). Présentation et interprétation de la méthode ELECTRE TRI pour affecter des zones dans des catégories de risque. Document du LAMSADE 124, LAMSADE, Université Paris Dauphine, Paris, France, (25 pages).

Yu, W. (1992). ELECTRE TRI: Aspects méthodologiques et manuel d'utilisation. Document du LAMSADE Nº 74, Université Paris-Dauphine, Paris, France.

Zopounidis, C., and Doumpos, M. (2000). Building additive utilities for multi-group hierarchical discrimination : The M.H.DIS method. *Optimization methods and Software* 14 : 219-240.

# APPENDIX A

## Proofs

*Proposition 2*

*Proof:*

Proposition 2.i: $xSB_k \Leftrightarrow$ There is $w \in B_{Lk}$ such that $xSw$ and there is no $z \in B_k$ fulfilling $zPx$. From Condition 3.vii, there is $y \in B_{Lh}$ ($k>h$) such that $wDy$ and hence $xSy$ (Condition 1.i). For each $z' \in B_{Lh}$ there is $z \in B_{Lk}$ such that $zDz'$ (Condition 3.viii). There could not be $z' \in B_{Lh}$ such that $z'Px$ because $z'Px$ and $zDz' \Rightarrow zPx$ (Propostion 1.ii) in contradiction with $xSB_k$. It follows that $xSB_h$ (Definition 1.a)

Proposition 2.i: $B_k Sx \Leftrightarrow$ There is $w \in B_{Uk}$ such that $wSx$ and there is no $z \in B_k$ fulfilling $xPz$. From Condition 3.v, there is $y \in B_{Uh}$ ($h>k$) such that $yDw$ and $ySx$ (Condition 1.i). For each $z' \in B_{Uh}$ there is $z \in B_{Uk}$ such that $z'Dz$ (Condition 3.vi). There could not be $z' \in B_{Uh}$ such that $xPz'$ because $xPz'$ and $z'Dz$

$\Rightarrow xPz$ (Proposition 1.i) in contradiction with $B_k Sx$. It follows that $B_h Sx$ (Definition 1.b).

## Proposition 3

*Proof:*

The proofs of Uniqueness, Independence, and Homogeneity are trivial, determined by the form the assignment rules were designed (see Definitions 2 and 3).

Monotonicity Property:

Primal procedure:

Let $C_{k*}$ be the category to which $x$ is assigned to. $x$ is assigned to $C_{k*} \Rightarrow$ There is $w \in B_{Lk*-1}$ such that $xSw$ and there is no $z \in B_{k*-1}$ fulfilling $zPx$ (Definition 1.a and Definition 2);

We have $yDx$ and $xSw$ ($w \in B_{Lk*-1}$) $\Rightarrow ySw$ from Condition 1.iii; (A)

There is no $z \in B_{k*-1}$ fulfilling $zPy$ since $zPy$ and $yDx \Rightarrow zPx$ (Proposition 1.i) in contradiction with $xSB_{k*-1}$ (B)

Combining (A), (B), and Definition 1.a we have $ySB_{k*-1}$; hence, according to Definition 2 $y$ should be assigned to $C_{k'}$ ($k' \geq k*$).

Dual procedure

Let $C_{k*}$ be the category to which $x$ belongs to. $x$ is assigned to $C_{k*} \Rightarrow$ There is $w \in B_{Uk*}$ such that $wSx$ and there is no $z \in B_{k*}$ fulfilling $xPz$ (Definition 1.b and Definition 3); Suppose that $y$ is assigned to $C_{k'}$ ($k' < k*$). From Definition 1.b and Definition 3, $y$ is assigned to $C_{k'} \Rightarrow$ There is $w \in B_{Uk'}$ such that $wSy$ and there is no $z \in B_{k'}$ fulfilling $yPz$;

Additionally, we have $wSy$ and $yDx \Rightarrow wSx$ for some $w \in B_{Uk'}$. (C)

There is no $z \in B_{k'}$ fulfilling $yPz \Rightarrow$ there is no $z \in B_{k'}$ fulfilling $xPz$, (D) since $xPz$ and $yDx \Rightarrow yPz$ (Proposition 1.ii); Combining (C) and (D) we have $B_{k'}Sx$ (Definition 1.b). According to Definition 3, $x$ should belong to class $C_k$ ($k \leq k'$). This contradicts the hypothesis.

## Proposition 4

*Proof:*

Primal procedure:

Suppose that $x \in B_{Lk'}$. Since $xSx$, from Conditions 2.iii, 3.i and Definition 1.a, we have $xSB_k$ with $k=k'$. Furthermore, $not(xSB_k)$ for $k>k'$ (Condition 3.ii). From the primal procedure, $x$ is assigned to $C_{k'+1}$. Suppose now that $x \in B_{Uk'}$. From Condition 3.i, $x$ does not fulfill $xSB_{k'}$ (Definition 1.a). But from Condition 3.iii, there is $y \in B_{Lk'-1}$ such that $xSy$. Combined with Condition 3.ii, we have $xSB_{k'-1}$. Hence, $x$ is assigned to $C_{k'}$.

Dual procedure:

Suppose that $x \in B_{Uk'}$. We have $B_{k'}Sx$ (Conditions 2.iv, 3.i and Definition 1.b). Also, $not(B_kSx)$ for $k<k'$ from Condition 3.ii. Then $x$ is assigned to $C_{k'}$. Suppose that $x \in B_{Lk'}$. We have $not(B_{k'}Sx)$ from Condition 3.i – Definition 1.b and $not(B_kSx)$ for $k<k'$ (Condition 3.ii). There is $y$ in $B_{Uk'+1}$ such that $ySx$ (Condition 3.iv). Then $B_{k'+1}Sx$ from Condition 3.ii and Definition 1.b. Hence, $x$ is assigned to $C_{k'+1}$.

***Proposition 5***

*Proof:*

Primal procedure:

    a. Merging operation between two consecutive categories.

        i. Consider that the action $x$ belongs to $C_h$, for $h > k+1$ (*i.e.*, $h-1>k$) before we proceed to a merging operation. Given the two propositions $not(xSB_h$ and $xSB_{h-1}$ are verified, after removing $B_k$, we obtain exactly the same situation; $x$ will belong to the same category as before once $B'_{h-1} = B_h$ and $C'_{h-1} = C_h$ for $h>k$.

        ii. Consider that the action $x$ belongs either to category $C_k$ or to category $C_{k+1}$ before we proceed to a merging operation. If before withdrawing $B_k$ $x$ was in category $C_{k+1}$, it naturally follows that $xSB_k$ and $not(xSB_h)$ for all $h>k$ (from Definition 2) and $xSB_{k-1}$ from Proposition 2.i. After we proceed to a merging operation, $x$ will belong to category $C'_k$, which is the category that combines the two categories $C_k$ and $C_{k+1}$. If before we proceed to a merging operation the action $x$ was in category $C_k$, from Definition 2 we have both $xSB_{k-1}$ and $not(xSB_h)$, for all $h>k-1$. After a merging operation, no one of the previous conditions changes; from Definition 2, the action $x$ is added to category $C'_k$, which is the category that combines both categories $C_k$ and $C_{k+1}$.

iii. Consider that action $x$ belongs to category $C_h$, for $h < k$ before we proceed to a merging operation. Evidently, after removing $B_k$, no one of the previous conditions changes; according to Definition 2, the action $x$ is classified into the same category as previously.

b. Splitting a class in two new consecutive ones.

i. Consider that the action $x$ belongs to $C_h$, for $h \geq k+1$ before we proceed to a splitting operation. According to Definition 2 and Definition 4(b), it naturally follows both conditions, $not(xSB'_{h+1})$ and $xSB'_h$, where $B'_h = B_{h-1}$ and $B'_{h+1} = B_h$. Thus $x$ will be assigned to category $C'_{h+1}$ (the old $C_h$).

ii. Consider that the action $x$ belongs to category $C_k$ before we proceed to a splitting operation. From Definition 2, we obtain both conditions, $not(xSB_k)$ and $xSB_{k-1}$. After inserting the set $B'_k$, according to Definition 4(b), we obtain both conditions, $not(xSB'_{k+1})$ and $xSB'_{k-1}$. From Definition 2 and $xSB'_k$, it follows that $x$ will be classified into category $C'_{k+1}$. If $x$ does not outrank $B'_k$, $x$ will be added to category $C'_k$. Thus, $x$ is will be added to one of the categories in which category $C_k$ was divided.

iii. Consider now that the action $x$ belongs to category $C_h$, for all $h < k$ before we proceed to a splitting operation. According to Proposition 2.i, $not(xSB_h)$ implies $not(xSB'_k)$. After splitting the category no one of the previous conditions change, given that $B'_h = B_h$, for all $h<k$, from Definition 4(b). Then, after we proceed to a splitting operation, the action $x$ will be added to the category $C'_h$ (which is the same category $C_h$), for all $h<k$.

Dual procedure:

The proof follows the same logic as above, but now using Proposition 2.ii and Definition 3 instead of Proposition 2.i and Definition 2.

The proof is complete.

*Proposition 6*

*Proof:*

Proposition 6.i: $xPB_k \Leftrightarrow$ There is $w \in B_k$ such that $xPw$ and there is no $z \in B_k$ fulfilling $zPx$. For each $z' \in B_h$ there is $z \in B_k$ fulfilling $zDz'$ (Condition 4.iv). There could not be $z' \in B_h$ such that $z'Px$ because $z'Px$ and $zDz' \Rightarrow zPx$ (Proposition 1.ii) in contradiction with $xPB_k$. It follows that $xPB_h$

(Definition 6.a).

Proposition 6.ii: $B_k Px \Leftrightarrow$ There is $w \in B_k$ such that $wPx$ and there is no $z \in B_k$ fulfilling $xPz$. For each $z' \in B_h$ there is $z \in B_k$ such that $z'Dz$ (Condition 4.v). There could not be $z' \in B_h$ such that $xPz'$ because $xPz'$ and $z'Dz \Rightarrow xPz$ (Proposition 1.i) in contradiction with $B_k Px$. It follows that $B_h Px$ (Definition 6.b).

*Proposition 7*

*Proof:*

The proofs of Uniqueness, Independence, and Homogeneity are obvious (see Proposition 3).

Monotonicity Property:

Dual procedure:

Let $C_{k*}$ be the category to which $x$ is assigned to. $x$ is assigned to $C_{k*} \Rightarrow$ There is $w \in B_{k*-1}$ such that $xPw$ and there is no $z \in B_{k*-1}$ fulfilling $zPx$ (Definition 6.a and Definition 8);

$\qquad yDx$ and $xPw$ $(w \in B_{Lk*-1}) \Rightarrow yPw$ from Proposition 1.ii; $\qquad$ (A)

There is no $z \in B_{k*-1}$ fulfilling $zPy$ since $zPy$ and $yDx \Rightarrow zPx$ (Proposition 1.i) in contradiction with $xPB_{k*-1}$ $\qquad$ (B)

Combining (A), (B), and Definition 7.a we have $yPB_{k*-1}$; hence, according to Definition 8 $y$ should belong to $C_{k'} (k' \geq k*)$.

The proof for the primal rule is omitted. It can be argued by the equivalence through the transposition operation.